\documentclass
	[a4paper,twocolumn,twoside,12pt,showkeys]
	{revtex4}

\usepackage{bm}					
\usepackage{newtxtext}
\usepackage{newtxmath}
\usepackage{color}					
\usepackage{graphicx}				
\usepackage{dcolumn}				
\usepackage[colorlinks=true, citebordercolor={0 0 0}, linkcolor=blue, citecolor=blue, urlcolor=blue]{hyperref}
\usepackage{setspace}
\usepackage{comment}

\newcommand{\figCapSkip}{\vspace{-4ex}}	
\makeatletter
\def\frontmatter@abstractwidth{0.9\textwidth}	
\makeatother

\usepackage{siunitx}
\usepackage{ulem}

\begin{document}


\newcommand{\By}{$\times$}
\newcommand{\SqrtBy}[2]{$\sqrt{#1}$\kern0.2ex$\times$\kern-0.2ex$\sqrt{#2}$}
\newcommand{\Degree}{$^\circ$}
\newcommand{\DegreeC}{$^\circ$C\,}
\newcommand{\Ohmcm}{$\Omega\cdot$cm}
\newcommand{\kay}{cm^{-1}}
\newcommand{\Sp}[1]{$\mathrm{sp}^{#1}$}
\newcommand{\EF}{$E_{\mathrm{F}}$}
\newcommand{\ADD}[1]{{#1}}
\renewcommand{\thesection}{\arabic{section}}
\renewcommand{\thesubsection}{\thesection.\arabic{subsection}}

\title{
Termination of Graphene Edges Created by Hydrogen and Deuterium Plasmas
}


\author{Taisuke Ochi}
\email[Corresponding author: ]{Taisuke.Ochi@anritsu.com}
\author{Masahiro Kamada}
\author{Takamoto Yokosawa}
\author{Kozo Mukai$^1$}
\author{Jun Yoshinobu$^1$}
\author{Tomohiro Matsui}
\email[Corresponding author: ]{Tomohiro.Matsui@anritsu.com}
\affiliation{%
Advanced Research Laboratory, Anritsu Corporation, 5-1-1, Onna, Atsugi, Kanagawa 243-8555, Japan
}

\affiliation{%
$^1$The Institute for Solid State Physics, The University of Tokyo, 5-1-5, Kashiwanoha, Kashiwa, Chiba, 277-8581, Japan
}


\begin{abstract}
\vspace*{1mm}

Edge engineering is important for both fundamental research and applications as the device size decreases to nanometer scale.
This is especially the case for graphene because a graphene edge shows totally different electronic properties depending on the atomic structure and the termination.
It has recently been shown that an atomically precise zigzag edge can be obtained by etching graphene and graphite using hydrogen (H) plasma.
However, edge termination had not been studied directly.
In this study, termination of edges created by H-plasma is studied by high-resolution electron energy loss spectroscopy to show that the edge is \Sp{2} bonded and the edge carbon atom is terminated by only one H atom.
This suggests that an ideal zigzag edge, which is not only atomically precise but also \Sp{2} bonding, can be obtained by H-plasma etching.
Etching of the graphite surface with plasma of a different isotope, deuterium (D), is also studied by scanning tunneling microscopy to show that D-plasma anisotropically etches graphite less efficiently, although it can make defects more efficiently, than H-plasma.
\end{abstract}



\maketitle
\newpage


\section{Introduction}

\vspace{-0.25\baselineskip}
Graphene, a two-dimensional sheet of carbon (C) atoms, is attracting vast interest and its bulk properties have been studied extensively since its discovery \ADD{\cite{Novoselov2004,Novoselov2005,Zhang2005,cao2018unconventional,toyama2022two,zhou2021superconductivity,yankowitz2019van,liu2022applications,ronen2021aharonov,deprez2021tunable}}.
One of the still-remaining frontiers of graphene research resides in its edges. 
Among two types of graphene edge structures, i.e., zigzag and  armchair, the biparticle symmetry is broken along the zigzag edge.
As a result, a flat band is expected to appear at the Fermi energy (\EF) around the K point.
Interestingly, this flat band can split due to spin polarization under electron-electron interaction, similarly to flat-band ferromagnetism \cite{fujita1996peculiar}.
Experimentally, the spin-unpolarized zigzag edge state has been observed on graphite surfaces by scanning tunneling microscopy and spectroscopy (STM/S) as a peak in the local density of state (LDOS) at around \EF, which is spatially localized only around the zigzag edge \cite{kobayashi2005observation,niimi2005scanning,niimi2006scanning}.
However, further study of the zigzag edge state including the spin-polarized state is limited because it is difficult to prepare a zigzag edge that is comparable to theoretical models.
Namely, the zigzag edge should be atomically precise and, at the same time, it should be terminated by only one hydrogen (H) atom to preserve \Sp{2} bonding of the honeycomb lattice along the edge.

\begin{figure} [tb]
  \begin{center}
    \includegraphics [
      width = 0.9\columnwidth
    ] {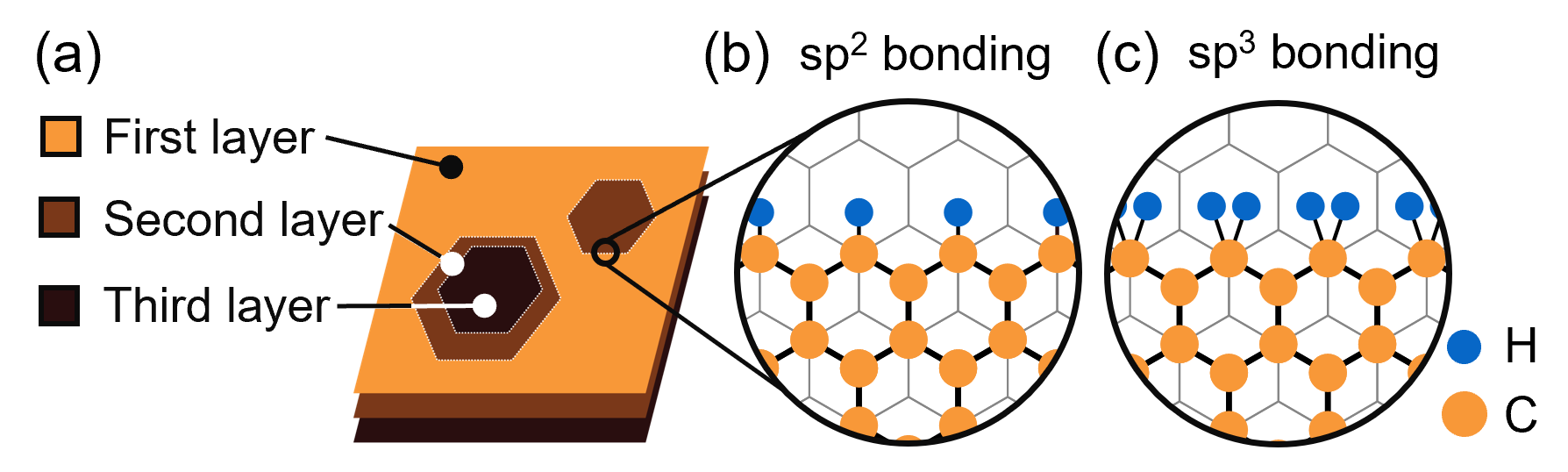}
  \end{center}\figCapSkip
  \caption {
    \label {fig_1}
(a) Illustration of a typical surface structure prepared by H- and D-plasma etching.
Some hexagonal nanopits are created layer by layer.
Atomic configuration of (b) \Sp{2}- and (c) \Sp{3}-bonded zigzag edges.
The edge C atom is terminated by one (two) H atom(s) for the \Sp{2} (\Sp{3})-bonded edge.
The hexagonal lattice of the second layer is also shown.
  }
\end{figure}

Recently, it has been found that hexagonal nanopits with atomically precise zigzag edges can be prepared on graphene and graphite surfaces by exposing them to remote H-plasma at high temperatures \cite{yang2010anisotropic,diankov2013extreme,hug2017anisotropic,matsui2019hexagonal,yokosawa2022nanoscale}.
Figure~\ref{fig_1}(a) illustrates a typical surface structure of graphite etched by H-plasma.
Even a LDOS suggestive of the spin-polarized state was observed for graphene nanoribbons terminated by parallel zigzag edges (z-GNRs) prepared using this H-plasma etching (HPE) technique.
However, termination of the edge is still unclear.
Considering that the edges are prepared by a chemical reaction with H, the edge C atoms are most probably terminated by H.
But the chemical bonding of the edge is not trivial.
It can be \Sp{2} bonded if the edge C atom is terminated by one H atom as illustrated in Figure~\ref{fig_1}(b), otherwise, it is \Sp{3} bonded if the edge C atom is terminated by two H atoms as shown in Figure~\ref{fig_1}(c).
Although an analysis of STM images together with a consideration of the chemical potential suggests that the edge is \Sp{2} bonded and terminated by one H atom \cite{zhang2013experimentally}, there are no direct observations that show edge termination unambiguously.


Edge termination of graphene nanoislands on Pt(100) and Pt(111) has been studied by high-resolution electron energy loss spectroscopy (HREELS) \cite{zecho1998hydrogen, dinger1999hydrogen}.
In these studies, nanoislands were prepared by thermal decomposition of hydrocarbons, and their edges were terminated by either H or deuterium (D) by exposing the nanoislands to thermally dissociated H or D.
They showed both stretching and bending modes of C-H ($\nu$CH and $\delta$CH) and C-D ($\nu$CD and $\delta$CD) clearly as isolated peaks in the spectra.
Edge termination was determined by comparing these stretching modes with those measured for submonolayer benzene and cyclohexane adsorbed on Pt(111) \cite{lehwald1978vibration,land1993hreels}.
Note here that benzene is composed of \Sp{2}-bonded C atoms, while cyclohexane is composed of \Sp{3}-bonded C atoms.
It was concluded that the edges of graphene nanoislands were \Sp{3} bonded and terminated by two H or D atoms, because the energies of the stretching modes were closer to those for cyclohexane than those for benzene.

Theoretically, on the other hand, the local vibrational density of states were calculated for GNRs with both zigzag and armchair edges terminated by only one H atoms, namely \Sp{2}-bonded edges \cite{vandescuren2008theoretical, islam2014effect}.
They showed an isolated peak at around \SI{2900}{\kay} and a bunch of peaks at lower energies than \SI{1800}{\kay}.
The authors suggested that the isolated peak at higher energy originated from the $\nu$CH at the edge, while the peaks at lower energies include edge-localized phonon modes and surface phonon modes.

In this work, termination of the edges created by the H- and D-plasma etching of graphite was studied by HREELS.
In contrast to the nanoislands on Pt surfaces studied previously \cite{zecho1998hydrogen,dinger1999hydrogen}, our samples were graphene nanopits on graphite surfaces. 
The $\nu$CH and the $\nu$CD were observed as isolated peaks at around \SI{2890}{\kay} and \SI{2143}{\kay}, respectively, suggesting that the edges prepared by HPE of graphite is \Sp{2} bonded and terminated by only one H atom.
In addition, it is also suggested that edge termination is robust even under ambient conditions.
Besides the HREELS study, the similarities and the differences between the etching by H- and D-plasma were also studied by STM. 

\section{Experimental}

The H- and D-plasma etched samples (HPE and DPE samples) were transferred from the etching chamber to the ultra-high vacuum (UHV) chamber for HREELS within several hours.
To eliminate unknown effects occurring by exposing the sample to the atmosphere, the edge of the DPE sample re-terminated by either H or D atoms in-situ in the UHV chamber (H- or D-modified sample) was also studied.
Here, the edge termination of the DPE sample was first removed by heating the sample up to $T \sim 1000$~K for 5 minutes (cleaned sample), and it was then exposed to H or D atoms, which were thermally dissociated from either $\mathrm{H}_2$ or $\mathrm{D}_2$ by a tungsten (W) filament ($\phi\ \SI{0.3}{\mm})$ at $T\sim\SI{1700}{\K}$ located about \SI{10}{\cm} away from the sample surface for 1 to 2 hours.
During this dissociation and re-termination process, the sample was cooled at $T\sim\SI{100}{\K}$ under $\mathrm{H}_2$ or $\mathrm{D}_2$ pressure of 5.0\By$10^{-6}$ \si{Pa}.
It was confirmed by STM that the surface structure was not changed by this procedure after the HREELS measurement.
Note that all the samples were baked at 400$-$600 \si{K} before the HREELS measurement to remove physisorbed species on the sample surface.

The HREELS measurements were performed for graphite ($\sim$\SI{12}{mm}\By\SI{12}{mm}\By$^{\mathrm{t}}$\SI{1}{mm}) rather than graphene to obtain sufficient signal from the edge.
The parameters for the H- and D-plasma etching were tuned to obtain as many hexagonal nanopits as possible.
By calculating the edge length from the STM images, one can expect that around 5.9\By$10^4$ and 8.4\By$10^4$ edge atoms/\si{\um^2} were prepared for the HPE and DPE samples, respectively, which is sufficient to obtain C-H and C-D vibrational signals by HREELS (ELS5000, LK Technologies). 

The HREELS spectra were obtained at $T\sim\SI{90}{K}$ with a primary electron energy of \SI{7}{eV} and an incident angle of \SI{60}{\degree} from the sample normal.
The energy resolution was \SI{48}{\kay} (\SI{6}{meV}) in a specular detection geometry.
The measurements were performed in an off-specular geometry, where the detection angle from the sample normal was \SI{50}{\degree} for the HPE, DPE and D-modified samples, while it was \SI{45}{\degree} for the H-modified and cleaned samples.
This difference in detection angle only affects the background intensity, and the C-H and C-D vibration signals are barely affected.

On the other hand, all STM images were obtained in constant current mode ($I=\SI{1.0}{\nA}$, $V=\SI{500}{\mV}$) under atmospheric conditions.



\section{Results and Discussion}

\subsection{Etching behavior of D-plasma}

\begin{figure} [tb]
  \begin{center}
    \includegraphics [
      width = 0.9\columnwidth
    ] {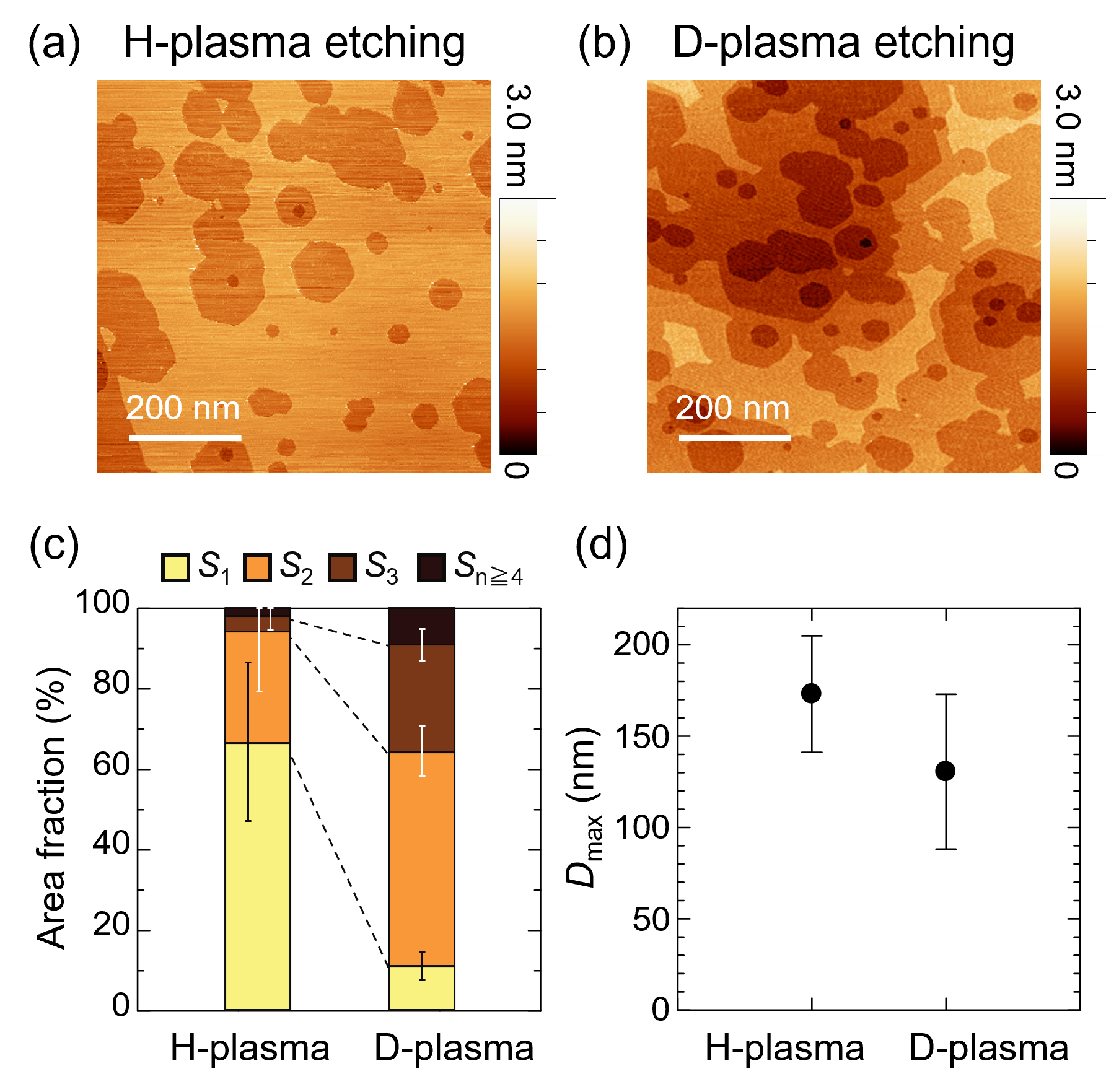}
  \end{center}\figCapSkip
  \caption {
    \label {fig_2}
    Comparison between H-plasma and D-plasma etchings. 
    (a) and (b) are STM images of graphite surfaces etched by H-plasma and D-plasma, respectively, with the same etching parameters. (c) and (d) are $S_{\mathrm{n}}$ and $D_{\mathrm{max}}$ extracted from the STM images, respectively.
  }
\end{figure}

First of all, the etching behavior of the D-plasma for the graphite surface was investigated.
Figure~\ref{fig_2}(a,\,b) shows the STM images of graphite surfaces etched by H- and D-plasma, respectively, with the same etching parameters.
For both samples, hexagonal nanopits are created and the etching behaviors are similar to each other.
This suggests that the chemical reaction of H- and D-plasma etching for graphite is identical.
However, the details are different.
For example, D-plasma etches the graphite surface more deeply than the H-plasma.
The difference is quantitatively analyzed by calculating the area fraction of the $n$-th layer from the surface ($S_{\mathrm{n}}$) and the maximum nanopit size of the first layer ($D_{\mathrm{max}}$) as shown in Figure~\ref{fig_2}(c,\,d), respectively.
The definitions of $S_{\mathrm{n}}$ and $D_{\mathrm{max}}$ follow Reference \cite{matsui2019hexagonal}.
Figure~\ref{fig_2}(c) clearly suggests that the larger surface area is etched away and that deeper layers appear by D-plasma etching than by H-plasma etching.
On the other hand, the $D_{\mathrm{max}}$ of the D-plasma etched sample is smaller just slightly than that of the H-plasma etched one as shown in Figure~\ref{fig_2}(d).
Considering that H-ions, such as $\mathrm{H}^{+}, \mathrm{H}_{2}^{+}, \mathrm{H}_{3}^{+}$, create surface defects and that H-radicals enlarge the defects into nanopits in HPE \cite{hug2017anisotropic,matsui2019hexagonal,yokosawa2022nanoscale}, the difference between H- and D-plasma etching suggests that D-ions create surface defects more efficiently than H-ions, while the effect of enlarging the defects by D-radicals is similar to or weaker than that by H-radicals, at least under these etching conditions.


\subsection{HREELS measurement}

\begin{figure} [tb]
  \begin{center}
    \includegraphics[
      width=0.9\columnwidth
    ] {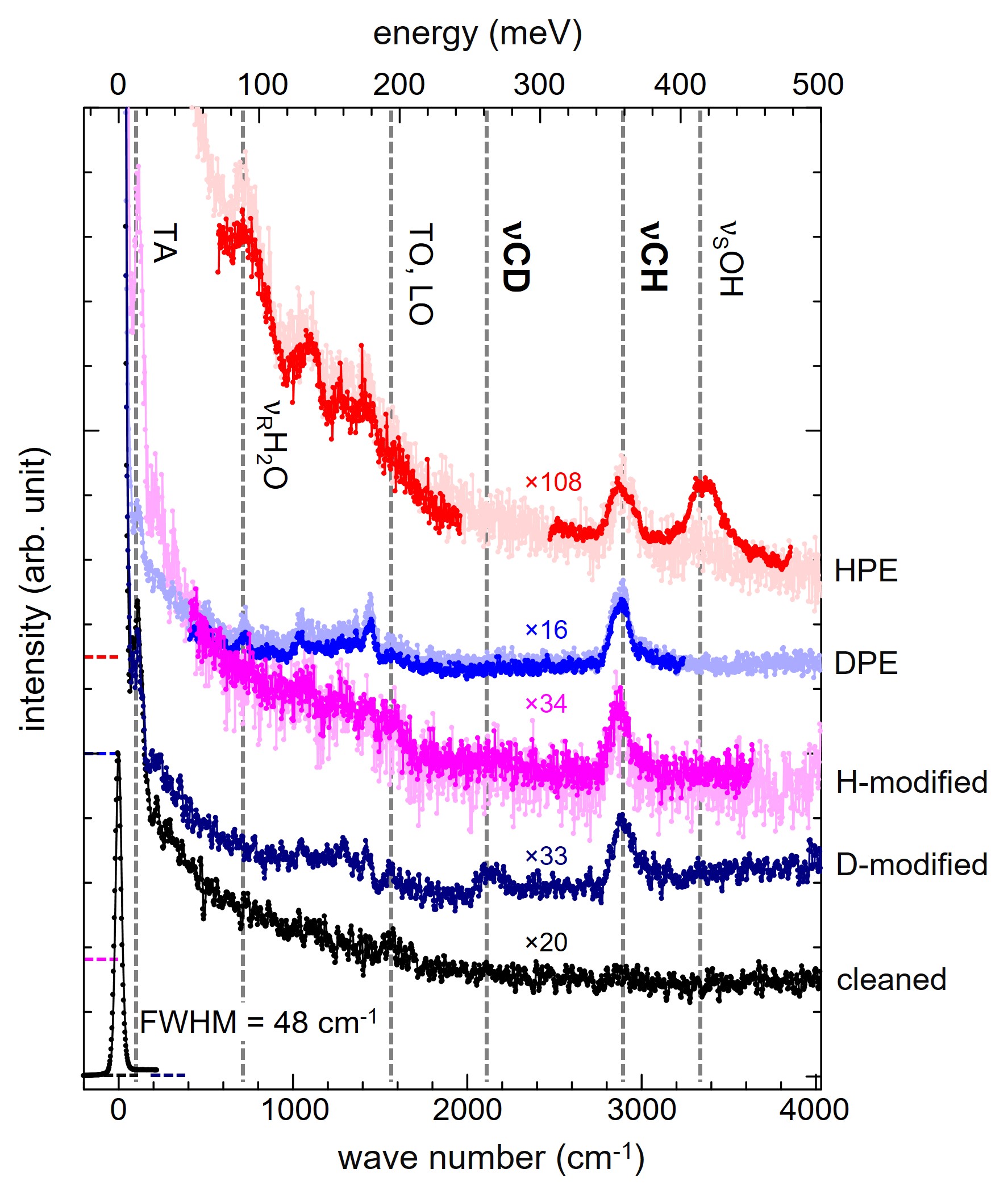}
  \end{center}\figCapSkip
  \begin{center}
    \caption{
      \label{fig_3}
HREELS spectra for the HPE, DPE, H-modified, D-modified and cleaned samples from top to bottom. 
The detection angle is \SI{50}{\degree} for the HPE, DPE and D-modified samples, while it is \SI{45}{\degree} for the other samples.
The spectra are shifted for clarity, and the horizontal dotted lines indicate the offset.
The vertical dotted lines show the energy losses of each vibration; 
$\nu_{\mathrm{S}}\mathrm{OH}$ and $\nu_{\mathrm{R}}\mathrm{H_{2}O}$ are $\mathrm{H_{2}O}$-related oscillations, while TA, LO and TO are the graphite surface phonon modes.
$\nu\mathrm{CH}$ and $\nu\mathrm{CD}$ are the C-H and C-D stretching modes, respectively.
    }
  \end{center}
\end{figure}

Figure~\ref{fig_3} shows the HREELS spectra for the HPE, DPE, H-modified, D-modified and cleaned samples.
Several electron energy loss peaks are observed for all the samples except the cleaned one.
For the HPE, DPE and H-modified samples, two spectra are overlaid, i.e., one with a wide energy range but with low energy resolution depicted by a light color and the other with narrow energy range but with high energy resolution depicted by a dark color.
Only for the HPE sample, peaks appear at around \SI{720}{\kay} and \SI{3360}{\kay}, the intensities of which change time by time.
They originate from the vibrational modes of $\mathrm{H_{2}O}$, namely the frustrated rotation (or libration) mode of $\mathrm{H_{2}O}$ ($\nu_{\mathrm{R}}\mathrm{H_{2}O}$) and the O-H stretching mode ($\nu_{\mathrm{S}}\mathrm{OH}$), respectively \cite{chakarov1995water}, due to the adsorption of water in the UHV chamber at low temperatures.
Note that such effect does not appear for the other samples.
For the cleaned sample, on the other hand, only energy losses related to the surface phonons of in-plane transverse acoustic (TA) mode at around \SI{110}{\kay} and in-plane transverse optical (TO) and longitudinal optical (LO) modes at around \SI{1580}{\kay} \cite{wirtz2004phonon, allouche2005hydrogen} are observed.
This indicates that the edge terminations are certainly removed by heating at $T\sim\SI{1000}{K}$.


For the samples other than the cleaned one, a bunch of peaks including bending mode and edge-localized phonon mode are observed at wave numbers lower than \SI{1600}{\kay}.
The spectra are too complicated to identify each peak and its origin.
On the other hand, an isolated peak is observed at energies higher than \SI{2000}{\kay} for all the samples except the cleaned one.
If the peaks are fitted by single gaussian, the peak positions for the HPE, DPE, H-modified and D-modified samples are at 2901, 2884, 2876 and 2902 \si{\kay}, respectively.
For the D-modified sample, another isolated peak is also observed at \SI{2143}{\kay}.
These two peaks at higher energies around \SI{2890}{\kay} and at \SI{2143}{\kay} can be the stretching modes. 
Considering that two of the samples, the H- and the D-modified samples, are prepared in-situ in UHV, these stretching modes can be related to the vibrations of the edge atoms rather than unknown contamination.
Since the energy ratio of these two peaks, which is about 1.35, is comparable to the one between $\nu\mathrm{CH}$ and $\nu\mathrm{CD}$ for edges of the graphene nanoislands on Pt surfaces \cite{dinger1999hydrogen, zecho1998hydrogen} and for submonolayer benzene \cite{lehwald1978vibration} and cyclohexane \cite{land1993hreels} on Pt(111), and is also comparable to the square root of the mass ratio between H and D, one can assign the peaks at around \SI{2890}{\kay} and \SI{2143}{\kay} to the $\nu$CH and $\nu$CD, respectively.


\begin{figure*} [tb]
  \begin{center}
    \includegraphics [
      width = 1.38\columnwidth
    ] {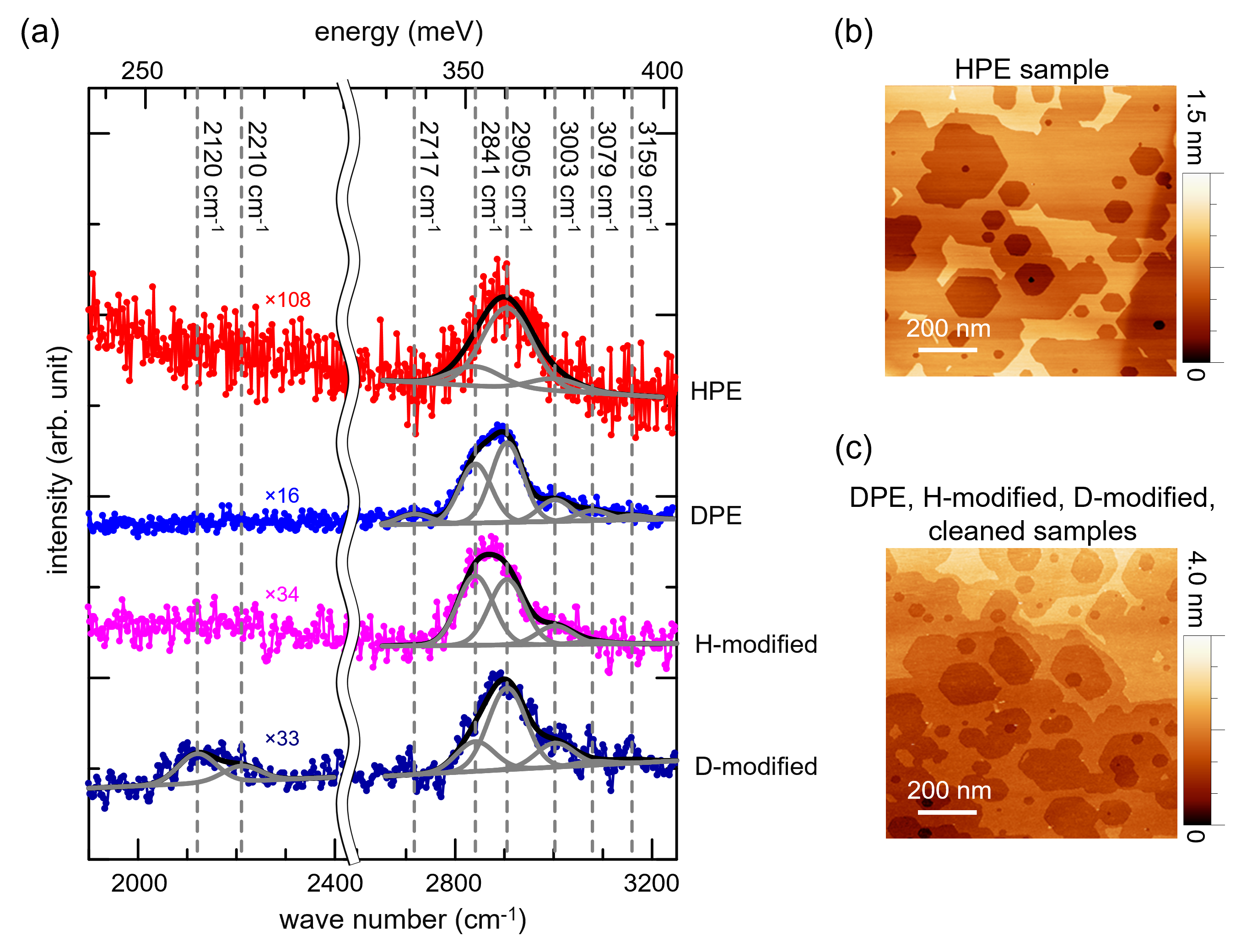}
  \end{center}\figCapSkip
  \caption {
    \label {fig_4}
(a) Gaussian fitting of $\nu\mathrm{CH}$ and $\nu\mathrm{CD}$ peaks.
Each gaussian position and the number of gaussians are determined using sparse modeling.
Each gaussian is shown in gray curves, while the sum of the component is shown in black.
(b,c) Typical topographic images of (b) the HPE sample and (c) the DPE, H-modified, D-modified and cleaned samples.
(b) is obtained on graphite which is etched by H-plasma simultaneously with the HPE sample for the HREELS.
On the other hand, (c) is exactly the surface studied by HREELS.
}
\end{figure*}


\ADD{Here}, the $\nu$CD appears weaker than the $\nu$CH for the D-modified sample and is even missing for the DPE sample.
This can be related to our sample preparation conditions.
In the D-modified sample, the $\nu$CD appears weaker than the $\nu$CH probably because D atoms stick on a graphite \ADD{edge} less than do H atoms, which are generated from the residual $\mathrm{H}_{2}$ gas in the UHV chamber.
The mass spectrum in the UHV chamber indicates that the $\mathrm{H}_{2}$ exist about \SI{1}{\%} of the $\mathrm{D}_{2}$ amount during the process.
\ADD{In fact,} the sticking probability of H atoms on graphite \ADD{surface} is \ADD{calculated to be} higher than that of D atoms at the sample preparation conditions in this study \cite{morisset2009isotopic}, namely the H and D atoms are thermally dissociated at $T\sim\SI{1700}{K}$ and are exposed to graphite at $T\sim$ \SI{100}{K}.
\ADD{It may also be the case on graphite edge.}
For the DPE sample, on the other hand, the $\nu$CD is missing probably because the D-plasma does not create hexagonal nanopits efficiently under the sample preparation conditions in this study.
Indeed, as discussed above, the anisotropic etching effect of the D-plasma is weaker than or similar to that of the H-plasma.
Therefore, the plasma of the residual $\mathrm{H_{2}}$ gas in the etching chamber rather than the D-plasma probably anisotropically etches the surface more efficiently under these etching conditions.
Note that the DPE sample is prepared in a chamber for the plasma etching which is made of a glass tube and evacuated by a rotary pump to $P\sim$ \SI{0.4}{Pa}.
This implies that more $\mathrm{H_{2}}$ gas remains in the etching chamber than in the UHV chamber for HREELS.
\ADD{However, further studies are desired to confirm these hypotheses and to fully understand these behaviors.}

\ADD{In any case,} it is confirmed unambiguously that the edges are terminated by H and/or D atoms.
To identify edge bonding, either \Sp{2} or \Sp{3}, we compare the stretching modes with previous measurements for graphene nanoislands\cite{dinger1999hydrogen, zecho1998hydrogen}, the counter-structure to the sample in this study, the nanopits.
The $\nu$CH at the edges of graphene nanoislands appeared at \SI{2700}{\kay} on Pt(111) and at \SI{2675}{\kay} on Pt(100), while the $\nu$CD was observed at \SI{1990}{\kay} on Pt(111) and at \SI{1940}{\kay} on Pt(100), which are significantly lower than those observed in this measurement.
Since the difference between the nanopits and the nanoislands is too large to explain by the difference in the structure and substrate, the bonding condition can be different between them.
Considering that the stretching mode for \Sp{2} bonding appears at around \SI{100}{\kay} higher energy than that for \Sp{3} bonding \cite{lehwald1978vibration,land1993hreels}, it can be expected that the stretching mode for the graphene nanopit (this study) is \Sp{2} bonding, while that for the graphene nanoisland (previous study) is \Sp{3} bonding.
It should be noted that, although the energies of $\nu$CH and $\nu$CD of the nanopits coincide with those of submonolayer cyclohexane, i.e., \Sp{3} bonding, on metallic substrates \cite{demuth1978ch, land1993hreels}, the assumption that the edge bonding of the nanopits is \Sp{3} is unlikely because the bonding of the edge of the nanoisland cannot be assigned if this is the case.
In addition, it is also difficult to expect that the edge bonding of a flat graphene shows the same energy within \SI{10}{\kay} as that of cyclohexane, the benzene ring of which has a steric structure.

Moreover, the $\nu$CH observed in this measurement at around \SI{2890}{\kay} agrees very well with the calculations of the local vibrational density of states for graphene nanoribbons with \Sp{2}-bonded edges, in which the $\nu$CH is calculated at around \SI{2900}{\kay} \cite{vandescuren2008theoretical, islam2014effect}.
This again suggests that the edges created by H-plasma etching are \Sp{2} bonded and terminated by only one H atom.
It is interesting to note that a bunch of peaks also appeared in these calculations similarly to our measurement.

Although the peaks at around \SI{2890}{\kay} and \SI{2143}{\kay} originate from the $\nu$CH and $\nu$CD of \Sp{2}-bonded edges, respectively, one can find that the peak is not a simple gaussian but contains detailed structure.
Therefore, each spectral peak is decomposed into several components by assuming that the peak is the sum of several gaussian peaks with the same FWHM (\SI{\sim50}{\kay} for the HPE sample and \SI{\sim35}{\kay} for the other samples) as shown in Figure~\ref{fig_4}(a).
The peak position and the number of peaks are determined using the LASSO (least absolute shrinkage and selection operator) technique of sparse modeling.
It is found that the $\nu$CH consists of four components, while the $\nu$CD consists of two components.
The fact that the stretching mode peak is composed of several gaussian peaks is possibly due to the imperfection of the edges and a somewhat complicated surface structure, since the vibrational energy loss can be shifted easily for about \SI{50}{\kay} by the different interaction with surrounding atoms \cite{dinger1999hydrogen, zecho1998hydrogen} and the displacement of H atoms from their equilibrium position \cite{islam2014effect}.
Therefore, although it is difficult to identify the origin of each component, it can be assumed that the two main components of $\nu$CH at \SI{2841}{\kay} and \SI{2905}{\kay} are related to the corner and the straight edge of the hexagonal nanopit, respectively.
Indeed, the ratio of the straight edge component at \SI{2905}{\kay} is larger among the other components for the HPE sample than the DPE sample, because the hexagonal nanopits are larger and have longer straight edges for the HPE sample than the DPE sample as can be seen in the STM images (Figure~\ref{fig_4}(b,c)).

\vspace{-0.25\baselineskip}

\section{Summary}

In this study, the differences between H- and D-plasma etching of graphite are studied by STM, and termination of the edges created by H- and D-plasma etching are studied by HREELS.
To avoid uncontrollable conditions such as surface contamination during the transfer of the sample from the etching chamber to the UHV chamber for HREELS, four kinds of graphite surfaces are studied throughout this measurement.
These are H- and D-plasma etched surfaces, and surfaces with H- and D-modified edges.
The last two samples are prepared based on the DPE sample in-situ in the UHV chamber for HREELS.

All the samples show a complicated peak structure at lower energies than \SI{1600}{\kay} and isolated peaks at around \SI{2890}{\kay} and \SI{2143}{\kay} in HREELS spectra. 
The isolated peaks are related to the $\nu$CH and $\nu$CD, suggesting that the edges are terminated by H and/or D atoms.
By comparing the stretching mode with previous studies on graphene nanoislands and theoretical calculations for graphene nanoribbons, it can be concluded that the edge created by H-plasma etching is \Sp{2} bonded and the edge C atom is terminated by only one H atom.
From the STM study of the etching behavior by H- and D-plasma together with this HREELS measurement, it is also found that D-plasma hardly etches graphite anisotropically although it can make defects more efficiently than H-plasma.
It is good to note, too, that the edge termination is robust even in air, since the C-H and C-D vibrations at edges can be observed by HREELS even after transferring the sample from the etching chamber to the UHV chamber.
From this study together with previous STM/S studies \cite{amend2018sts}, it can be concluded that not only atomically precise but also \Sp{2}-bonded zigzag edges can be obtained by the H-plasma etching of graphite and graphene.
Thus, this sample preparation technique provides an opportunity to study the exotic properties of graphene zigzag edges such as a spin-polarized zigzag edge state.

\begin{acknowledgments}
HREELS measurement was carried out by the joint research between Anritsu Corporation and The Institute for Solid State Physics, The University of Tokyo.
The authors acknowledge the free-to-use software, WSxM\cite{horcas2007wsxm} and Imaje-J Fiji\cite{schindelin2012fiji}.
\end{acknowledgments}

\bibliographystyle{elsarticle-num}
\bibliography{ref}

\clearpage

\setcounter{figure}{0}
\renewcommand{\thefigure}{S\arabic{figure}}

\section{Supplementary}

\subsection{Effect of the sample annealing}

\begin{figure} [t]
  \begin{center}
    \includegraphics [
      width = 0.9\columnwidth
    ] {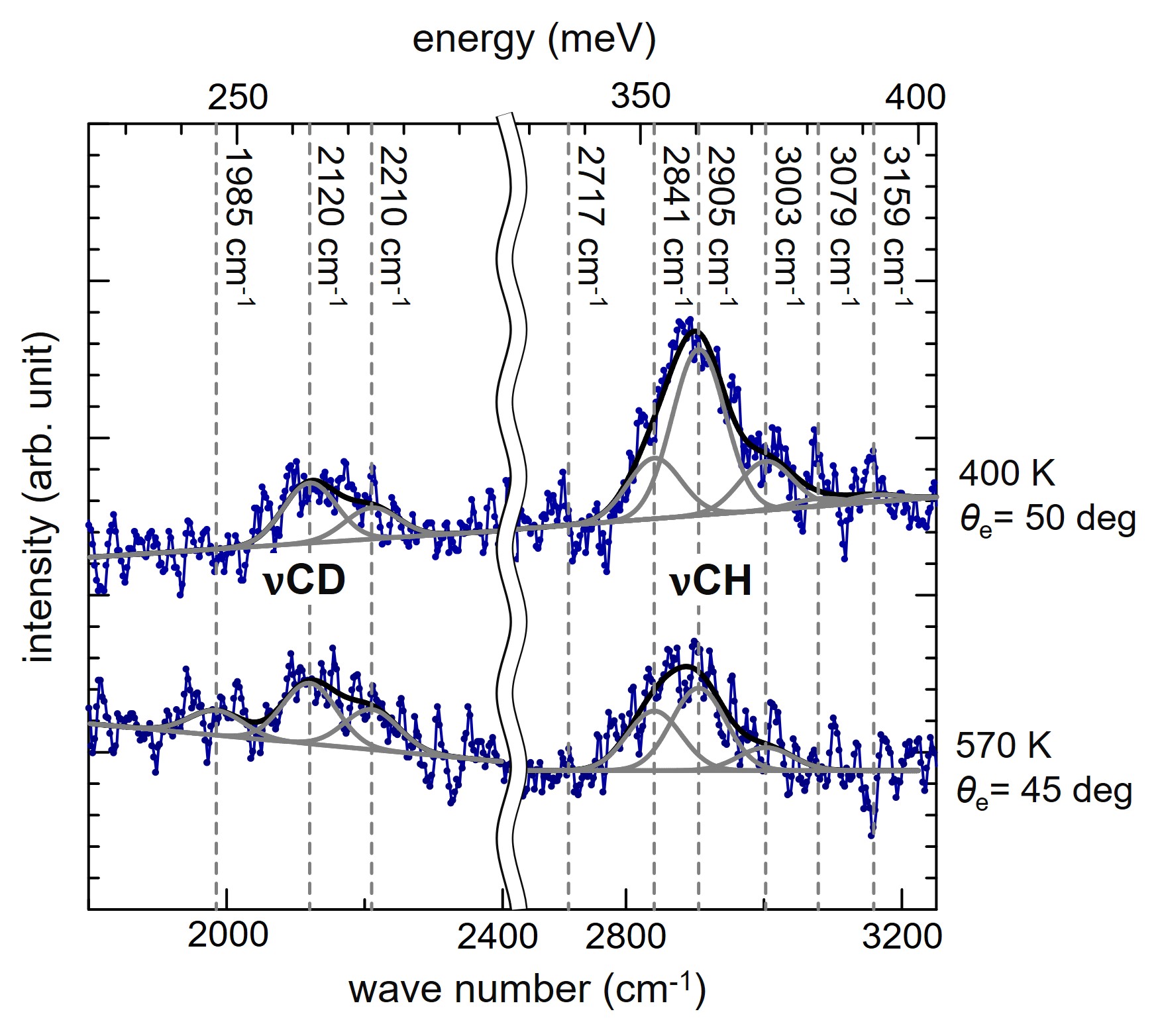}
  \end{center}\figCapSkip
  \caption {
  HREELS spectra for the D-modified sample after annealing at $T =$ \SI{400}{K}  (upper) and \SI{570}{K} (lower).
  The detection angle ($\theta_\mathrm{e}$) are \SI{50}{\degree} and \SI{45}{\degree}, respectively.
  The spectral peaks are decomposed into several gaussian peaks using a sparse modeling technique.
    \label {fig_S1}
  }
\end{figure}

The effect of the sample annealing is studied for the D-modified sample.
The spectra around the $\nu$CH and the $\nu$CD after the annealing at $T =$ \SI{400}{K} and \SI{570}{K} are shown in Figure~\ref{fig_S1}.
It is found that the edge termination is robust up to $T =$ \SI{400}{K}.
Intuitively, just the peak intensity decreases as the edge terminating H/D atoms are removed by increasing the anneal temperature.
However, when it is annealed at $T =$ \SI{570}{K}, not only the peak intensity but also the peak structure are modified.
Therefore, each peak is decomposed into several gaussian peaks using a sparse modeling technique (The details are the same as written in the main text).

For the $\nu$CH, only the peak intensity at \SI{2905}{\kay} decreases by annealing at $T =$ \SI{570}{K}.
Considering that the elemental peak at \SI{2905}{\kay} is related to the $\nu$CH on the straight edge of nanopit, while the elemental peak at \SI{2841}{\kay} is related to the one at the corner edge, this result suggests that the edge termination along the straight edge might be easier to be removed by heating than the termination at the corner edge.

For the $\nu$CD, on the other hand, an additional peak appears at \SI{1985}{\kay}.
Interestingly, this peak energy coincides with the stretching mode of \Sp{3}-bonded C-D at the graphene nanoisland edge on Pt surfaces \cite{dinger1999hydrogen, zecho1998hydrogen}.
Therefore, this fact implies that the edge termination changes from \Sp{2} to \Sp{3} by annealing.

\end{document}